\documentclass[lettersize,journal]{IEEEtran}
\usepackage{amsmath,amsfonts}
\usepackage{algorithmic}
\usepackage{algorithm}
\usepackage{array}
\usepackage[caption=false,font=normalsize,labelfont=sf,textfont=sf]{subfig}
\usepackage{textcomp}
\usepackage{stfloats}
\usepackage{url}
\usepackage{verbatim}
\usepackage{graphicx}
\usepackage{cite}
\begin{document}

\title{On the Reception Process of Molecular Communication-Based Drug Delivery}
\author{Roya Paridar, \IEEEmembership{Member, IEEE}, Nader Mokari, \IEEEmembership{Member, IEEE}, Eduard Jorswieck, \IEEEmembership{Fellow, IEEE}, and Mohammad Reza Javan, \IEEEmembership{Member, IEEE}
	\thanks{R. Paridar and N. Mokari are with the Department of Biomedical Engineering, Tarbiat Modares University, Tehran, Iran (e-mails: roya.paridar@modares.ac.ir, nader.mokari@modares.ac.ir).}
	\thanks{M. Javan is with Department of Electrical and Robotic Engineering, Shahrood University of Technology, Shahrood, Iran (e-mail: javan@shahroodut.ac.ir).}
	\thanks{E. Jorswieck is with institute of Communications Systems, TU Braunschweig, Germany. (e-mail: jorswieck@ifn.ing.tu-bs.de).}}
\maketitle
\begin{abstract}
	One of the important applications of molecular communication is the targeted drug delivery process in which the drug molecules are released toward the target (receiver) in a way that the side effects are minimized in the human body. As the total number of released molecules increases, the receiver cannot receive all of the transmitted drug molecules. Therefore, the molecules would be accumulated in the system which results in side effects in the body. In order to diagnose the appropriate transmission rate of the drug molecules, it is important to investigate the reception process of the receiver. In this paper, a reception model is studied using queuing theory. In the proposed model, the rejection rate of the drug molecules due to different reasons, such as random movement of the molecules, as well as the rejection rate due to active receptors are taken into account. Moreover, an interval consisting of the lower and upper bounds for the number of released molecules is presented based on the proposed model in order to determine the range of allowable dosage of the drug molecules. It is shown that the queuing theory can be successfully employed in accurate modeling of the reception process of the receiver in drug delivery applications.
\end{abstract}


\section{Introduction} \label{sec:introduction}
\IEEEPARstart{M}{olecular} communication (MC) is an emerging communication technique in which molecules are used to transfer information among nanomachines \cite{nakano2014molecular,akyildiz2008nanonetworks}. Each nanomachine is a nano-scale device which can perform a simple task in MC. Cooperation of nanomachines leads to performing a complex task \cite{nakano2012molecular}. This kind of communication can be found in biological systems such as communication between neurons in the human body nervous system \cite{veletic2019synaptic,balevi2013physical}. MC has a wide range of applications in the medical field which can be categorized into two main parts: diagnosis and treatment of disease \cite{felicetti2016applications}. Targeted drug delivery is one of the branches of MC applications in the treatment category which attracts a lot of recent attention among the researchers \cite{zhao2021release,chude2017molecular,islam2021molecular}. In targeted drug delivery, the main goal is to manage the release rate of the drug molecules in a way that the diagnostic target (e.g., the cancerous cell) is attacked by the drug molecules while the side effects to the other parts of the body are minimized \cite{khoshfekr2019drug}. In the targeted drug delivery system, the drug molecules are injected toward the target site, and the released molecules are freely diffused in the environment until they reach the target. This type of drug delivery system is known as the local drug delivery system \cite{salehi2018lifetime}, which is the main focus of this paper. It should be noted that all of the released drug molecules would not reach the target; some of them would be lost due to some reasons such as reaction with other molecules and the non-ideal performance of the receptors. On the other hand, it should be noted that if the concentration of the drug molecules exceeds the acceptable threshold, other parts of the body would be affected by the side effects of the drug molecules. Therefore, it is important to release the drug molecules in an appropriate rate in order to keep the drug dosage at the satisfied level during the delivery procedure \cite{siepmann2012fundamentals}. Since optimizing the release rate of the drug molecules greatly depends on the receive rate of the target, it is necessary to first examine the behavior of the target, or equivalently, the receiver part of the MC system. Therefore, the main focus would be on the receiver in the following.\\
\indent The MC system consists of three main parts: the transmitter, the channel, and the receiver. One of the important issues in MC is the receiver structure. Generally, the receiver structure is categorized into three main categories \cite{sun2020expected}:
\begin{enumerate}
	\item Fully cover receiver in which it is assumed that infinite receptors exist on the surface of the receiver. Therefore, the messenger molecule would be counted once it reaches the receiver.
	\item Partially cover receiver in which it is assumed that a part of the receiver consists of infinite receptors. Therefore, the messenger molecule would be counted only if it reaches the part of the receiver which consists of the receptors.
	\item Finite receptors in which it is assumed that the receiver consists of a finite number of receptors distributed uniformly on the surface of the receiver. Therefore, the messenger molecule would be counted only if it reaches the receptor, not other parts of the receiver. 
\end{enumerate}

Finite number of the receptors in the receiver part is an important issue in the receiver structure which is necessary to be considered; due to the limited number of the receptors, the messenger molecules (ligands) are accumulated in the environment and result in some problems in the system. In particular in the drug delivery application, this issue leads to the side effects in the human body. In addition to the finite number of receptors, the interaction between the ligand and the receptor is also important to be considered in order to achieve a model which is closer to reality. More precisely, the following parameters are important to be considered which leads to a more accurate model:
\begin{itemize}
	\item The ligand-receptor binding rate which results in the receptor to be in the active mode.
	\item The ligand-receptor unbinding rate which results in the receptor to be in the inactive mode.
	\item The rejection rate, which is defined as the case where the ligand is lost after a while without any binding process.
\end{itemize}

\noindent In many studies, the above parameters have been considered in the system model; the binding and unbinding rates of the ligand-receptor complex are taken into account. The rejection rate of the ligand due to the activation of the receptors is also considered. However, it should be noted that it is possible for a ligand to be rejected even if the receptors are in the inactive mode. In other words, it is possible for a molecule to enter the system. However, it leaves the system after a while without any binging process. This case would occur due to several reasons including the random movement of the ligands, the lifetime of the ligands, and the non-ideal performance of the receptors. In this paper, a model for the reception process of the receiver is proposed in which the rejection rate of the ligands in the case that the receptors are in the inactive mode is also considered. Following the proposed reception model, an interval consisting of the lower and upper bounds is proposed to determine the allowable transmission rate of the drug molecules.
\subsection{Related Works}
In many studies, the receiver is assumed to be a fully cover receiver \cite{jamali2017design,jamali2018constant,kuran2011modulation,rouzegar2017channel}. In these models, the finite number of the receptors is not taken into consideration. Moreover, the interaction between a receptor and a messenger molecule is ignored. In some cases, such as some processing performed by cells, having a fully cover receiver is an appropriate assumption \cite{andaloussi2013extracellular}. Also, in some cases where the main focus of the study is on the other parts of the MC system, such as designing a modulation technique \cite{kuran2011modulation,gursoy2019pulse}, the receiver is usually considered to be fully cover. However, in order to study the behavior of the receiver more accurately, it is necessary to take some conditions into account, such as the limited number of receptors and the interaction between the ligand and the receptor. A list of related works contains the following. in \cite{ahmadzadeh2016comprehensive}, the interaction between the receptors and the ligands is considered. In this model, it is assumed that a receptor can interact with multiple ligands simultaneously. Moreover, the environment is considered to be unbounded. This model is generalized to a model with a bounded environment in \cite{al2018modeling}. Moreover, the binding and unbinding rates of the ligand-receptor are considered in this model. Note that in the model proposed in \cite{al2018modeling}, no limitation is considered for the number of the receptors. In \cite{yuan2018performance}, the binding and unbinding rates of the ligand-receptor are also considered, and the number of received ligands is modeled using the Poisson distribution. In \cite{sun2020expected}, it is assumed that the number of the receptors is finite. Then, the considered model is approximated with the fully cover receiver to solve the problem. In \cite{kuscu2018maximum}, the finite receptor model is considered, and a maximum likelihood detection-based method is presented using the ligand-receptor unbinding duration. In the study conducted in \cite{kuscu2018maximum}, only one type of ligands and receptors is assumed. The mentioned model is generalized to multiple types of the receptors and ligands in \cite{kuscu2019channel} and a detection method is presented in order to detect the concentration of each type of the ligands. In \cite{aminian2015capacity}, the blocking effect of the receptors caused by other types of molecules is considered in the finite receptor model. In \cite{femminella2015molecular}, with the consideration of the finite receptor model, the rejection rate of the ligand by the receptor is taken into account. Note that in this study, it is assumed that the rejection rate occurs due to the activation of all of the receptors. This problem is further investigated in \cite{felicetti2016simple} and it is shown that the positions of the receptors on the receiver surface can be approximated to be concentrated in a single point. Considering the finite receptor model, in \cite{zhao2020release} the probability distribution of the active receptors is obtained in steady-state using the queuing theory principles, and the minimum release rate of the ligands is optimized at the transmitter part.\\
\indent The rejection rate of the ligands in the case where the receptors are in the inactive mode is not considered in the recent studies. Also, the rejection process due to active receptors, which is investigated in \cite{femminella2015molecular,felicetti2016simple}, affects the enter rate of the molecules; it is assumed that when the molecule, which is entering the system, faces an active receptor, it is rejected from the system. However, in the proposed reception model, the rejection rate, due to active receptors or random movement of the molecules is investigated in the exit rate of the molecules from the system. Moreover, in optimizing the release rate of the drug molecules, a lower bound is determined in the previous studies, as mentioned before. However, the upper bound of the release rate is not specified to the best of our knowledge. In this study, in addition to the lower bound obtained according to the proposed model, an upper bound for the number of releasing drug molecules is also proposed.
\subsection{Our Main Contribution}
In this paper, a finite receptor model is considered. Also, it is assumed that the ligands are freely diffused in the environment. The capacity of the environment is considered to be bounded; it is assumed that there is a reception space around the receiver where a limited number of ligands can be placed in this space \cite{meng2014receiver}. The system model is described using the queuing theory \cite{papoulis1989probability}. In the proposed model, the rejection rate of the ligands is also taken into account. Note that despite of \cite{femminella2015molecular,felicetti2016simple}, in the considered model, it is possible for a ligand to be rejected even if the inactive receptors exist in the system. In other words, due to the random movement of the ligands, it is possible for a ligand to enter the environment, walk randomly for a while, and then leave the system without any binding process. Another difference between the proposed reception model and \cite{femminella2015molecular,felicetti2016simple}, is that the rejection rate is investigated in the exit rate of the molecules; it means that the enter rate of the molecules is considered to be constant in the proposed model. After describing the proposed model, an interval consisting of the lower and upper bounds is expressed for the release rate of the drug molecules. Our main contributions are listed as below:
\begin{itemize}
	\item The rejection rate is considered in the model. More precisely, it is assumed that the ligands in the reception space can be rejected even if the receptors are inactive. Note that the rejection rate is investigated in the exit rate of the molecules from the system.
	\item The reaction rates (binding and rejection) of the ligands in the model are considered to change at each step. More precisely, each state in the considered queuing model equals the number of existing molecules in the system (or equivalently, the concentration of the molecules), and the reaction rates are considered to be state-dependent. Based on queuing theory, a mathematical model is obtained for each of these parameters.
	\item Mathematical analysis is performed to obtain the lower bound corresponding to the proposed model for the release rate of the drug molecules. In addition to the lower bound, an upper bound is also obtained. Consequently, an interval, consisting of a lower and an upper bound, is expressed with the aim of managing the release rate of the drug molecules in drug delivery application.
\end{itemize}
\begin{figure}[t]
	\centering
	\includegraphics[width=0.5\textwidth]{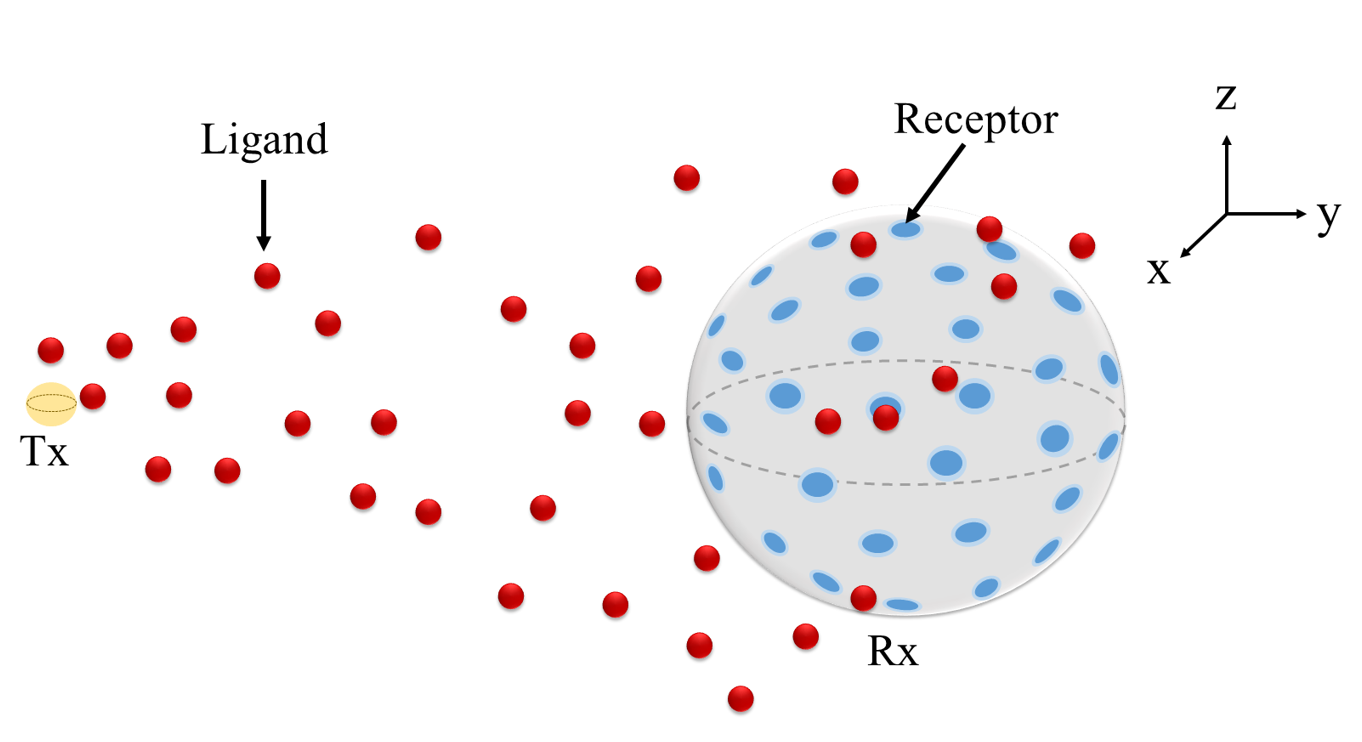}
	\caption{
		Schematic of the considered system. `'Tx" denotes the transmitter, and `'Rx" denotes the receiver.}
	\label{fig:one_type}
\end{figure}
The paper is organized as follow. In Section \ref{systemmodel}, a brief explanation about the system model is presented. In Section \ref{RPmod}, the proposed reception process is discussed, and the mathematical model is expressed for enter and exit rates of the ligands. In Section \ref{Bounds} the upper bound for the number of released molecules is presented in addition to the lower bound. Numerical simulations illustrate the usefulness of the proposed model, and also the desired bounds for releasing the drug molecules are presented in Section \ref{results}. Finally, the conclusion is presented in Section \ref{conclusion}.
\section{System Model} \label{systemmodel}
In this paper, a 3D MC system is considered where the ligands are freely diffused in the environment. Moreover, it is assumed that the number of the receptors at the receiver side is finite and equals $N_r$. The receiver can receive the ligand molecule once the molecule reaches the receptor, not other parts of the receiver. The schematic of the considered system is depicted in Fig. \ref{fig:one_type}. The transmitter, which is defined as the location where the drug molecules are released, is considered to be a point source. In drug delivery applications, the continuous transmission is performed with a specific transmission rate in order to affect the target side. Binary concentration shift-keying (BCSK) modulation technique is performed. The BCSK modulation technique is formulated as below \cite{atakan2007information}:
\begin{flalign}
Q_{TX}(t)=\begin{cases}
Q  & \quad \text{for } jT_s<t<(j+1)T_s,\\
0  & \quad \text{otherwise, }
\end{cases}
\label{eq:BCSK}
\end{flalign}
where $j\in\mathbb{N}_0$, and $Q_{TX}(t)$ denotes the concentration of ligands emitted from the transmitter at time instance $t$. According to this modulation technique, a constant number of ligand molecules $Q$ is transmitted during the pulse duration $T_s$. Based on this assumption, the reception model is discussed in the following.
\begin{figure*}[htbp]
	\centering
	\includegraphics[width=0.8\textwidth]{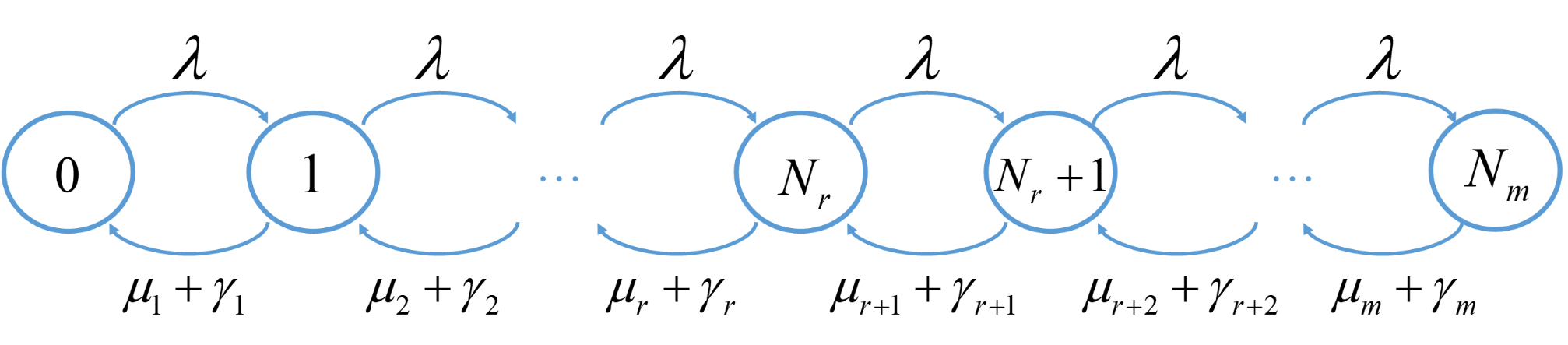}
	\caption{
		Schematic of the $M/M/N_r/N_m$ system where $N_m>N_r$.}
	\label{fig:birth_death}
\end{figure*}
\section{The Proposed Model based on Queuing Theory}\label{RPmod}
\subsection{Reception Process}
To analyze the reception procedure, the $M/M/N_r/N_m$ queuing system is considered where the molecules arrive at the reception space according to the Poisson distribution with the arrival rate of $\lambda$. It is assumed that at most $N_m$ number of molecules can be placed in the reception space. The ligand-receptor complex is constructed once the molecule binds to the receptor. It is assumed that the receptor cannot interact with other molecules as long as the ligand-receptor binding process is retained. The time interval between two consecutive unbinding processes, which is obviously the result of the binding process, can be modeled using the exponential distribution with the parameter $\mu$ \cite{papoulis1989probability,femminella2015molecular}. This parameter ($\mu$) can be interpreted as the service rate. In other words, the time required for constructing a ligand-receptor complex, staying in the active mode and finally, re-constructing another ligand-receptor complex is modeled as $\exp(\mu)$. The expected value of the exponential distribution equals $1/\mu$. This demonstrates that as the parameter $\mu$ increases, the time interval between two consecutive unbinding process would be decreased. When a molecule enters the environment, two scenarios can occur: the molecule binds to the receptor, or it leaves the environment without any binding process. The second scenario, which is known as the \textbf{rejection process}, is also modeled using the exponential distribution with the rejection rate of $\gamma$. More precisely, consider that one molecule leaves the environment without any binding process (rejection process). The time duration it takes for another molecule in the system to be rejected after the last rejection process, is modeled using the exponential distribution with the parameter $\gamma$ (similar to unbinding process). The time interval between two consecutive rejection processes is modeled as $\exp(\gamma)$.
This process is known as `'birth-death" process where entering a molecule equals the birth process, and leaving a molecule equals the death process. Based on the entering or leaving of a molecule, the state of the system is constantly changing. According to the above explanations, going from one state to the other state is modeled using the exponential distribution with the following probability density:
\begin{flalign}
\lambda_{n,n+1}&=\lambda,\nonumber\\
\lambda_{n,n-1}&=\mu_n+\gamma_n,
\label{eq:prob-density}
\end{flalign}
where $n$ denotes the state of the system. In the considered system, the $n^{th}$ state is interpreted as the presence of $n$ molecules in the reception space, or equivalently, the concentration of the molecules in the system. Note that the maximum number of molecules that can be placed in the reception space is $N_m$. Therefore, the expressed probability density is true for $n\leq N_m$. Also, note that $\lambda_{n,n+1}$ demonstrates the rate in which the system goes from state $n$ to the state $n+1$, or equivalently, the birth process. Similarly, $\lambda_{n,n-1}$ shows the rate in which the system goes from state $n$ to the state $n-1$, or equivalently, the death process. It is assumed that the discussed process has a constant birth rate. In other words, the molecules enter the reception area with a constant rate. However, the death rate depends on the status. Fig. \ref{fig:birth_death} shows the state transition diagram of the explained birth-death process. In the following, how to obtain the rates is discussed in detail.
\subsection{Enter and Exit Rates}
\begin{figure}[b]
	\centering
	\includegraphics[width=0.5\textwidth]{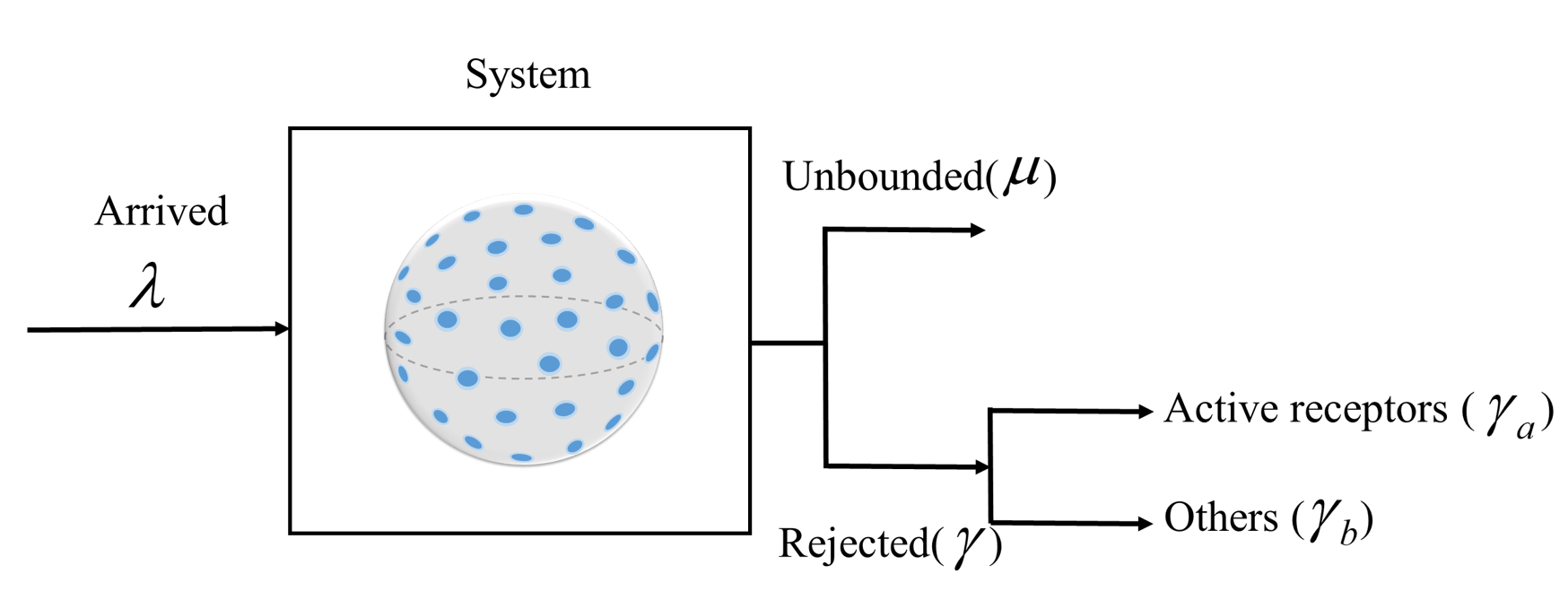}
	\caption{
		Schematic of the entering and exit rates of the molecule to the system.}
	\label{fig:Rates}
\end{figure}

In order to analyze the enter and exit rates of the molecules in the system, consider Fig. \ref{fig:Rates}. According to this figure, and also according to the explanations presented above, it can be concluded that the molecules enter the system with the constant arrival rate $\lambda$. Also, after spending a time duration, they leave the system. Leaving the system occurs due to the unbinding process, or the rejection process. The former reason is shown as `'Unbounded" in Fig. \ref{fig:Rates}. Also, the latter is shown as `'Rejected". As it can be seen from the figure, the rejection process is divided into two categories; 1- rejection due to the fact that the receptors are in the active mode, and 2- the rejection due to other reasons, such as random movement of the molecules.
\subsubsection{Enter Rate}
As mentioned before, the considered system is 3D. The impulse response of the diffusion equation in a 3D system is calculated as below \cite{zhao2020release}:
\begin{flalign}
h(t,r)=\frac{Q}{\left(4\pi Dt\right)^{3/2}}\exp\left(-\frac{r^2}{4Dt}\right),
\label{eq:impulse}
\end{flalign}
where $D$ denotes the diffusion coefficient, $t$ denotes the time, and $r$ represents the distance between the transmitter and the center of the receiver. Consider the continuous transmission, which is performed in drug delivery application, the drug molecules are released with the rate of $Q/\Delta t$. This continuous transmission can be considered as the transmission of a large number of pulses, starting at $t=0$. The concentration of the drug molecules at time $t$ at the distance $r$ from the transmitter is written as below \cite{zhao2020release}:
\begin{flalign}
c(r,t)&=\sum_{i=0}^{t/\Delta t}h(r,t-i\Delta t)\nonumber\\
&=\frac{1}{\Delta t}\sum_{i=0}^{t/\Delta t}h(r,t-i\Delta t)\Delta t\nonumber\\
&\approx\frac{1}{\Delta t}\int_{0}^{t}h(r,\tau)d\tau.
\label{eq:concentraion}
\end{flalign}
Inserting (\ref{eq:impulse}) into (\ref{eq:concentraion}), we have:
\begin{flalign}
c(r,t)\approx\frac{Q}{\Delta t4\pi Dr}\text{erfc}\left(\frac{r}{(4Dt)^{1/2}}\right),
\label{eq:concentraion2}
\end{flalign}
where $\text{erfc}(.)$ denotes the complementary error function \cite{farsad2016comprehensive}. In the steady-state condition, where $t\rightarrow\infty$, the concentration of the drug molecules in the system is obtained as $\frac{Q}{4\pi Dr\Delta t}$. This value is equivalent to the enter rate of the molecules by considering the distance between the transmitter and the center of the target $R$. More precisely, we have:
\begin{flalign}
\lambda=\frac{Q}{4\pi DR\Delta t}.
\label{eq:entering_rate}
\end{flalign}
As expected, the enter rate depends on the total number of transmitted molecules and the distance between the transmitter and the receiver; as the total number of transmitted molecules increases, the enter rate of the molecules is increased. Also, the enter rate is decreased by increasing the distance between the transmitter and the receiver.
\subsubsection{Exit Rate Due to Rejection Process}
\begin{figure}[t]
	\centering
	\includegraphics[width=0.35\textwidth]{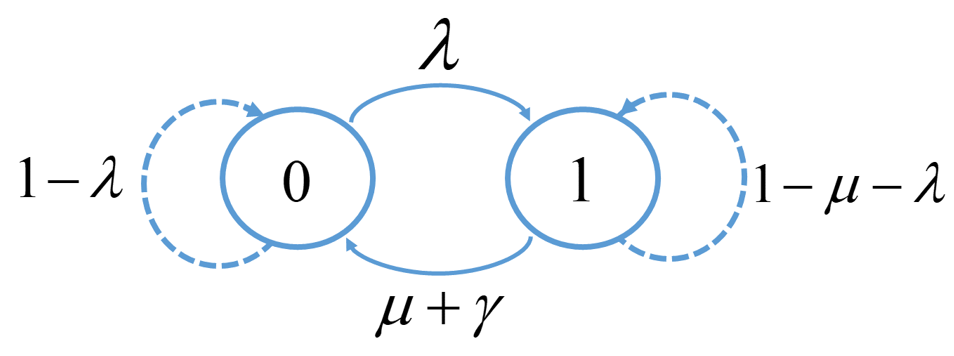}
	\caption{
		Schematic of the $M/M/1/1$ system.}
	\label{fig:single}
\end{figure}

The exit rate of the molecule is equivalent to $\lambda P_{\text{rej}}$, where $P_{\text{rej}}$ is the probability of rejection. In the considered system, the enter rate of the molecules equals the total exit rate of the molecules. In other words, considering Fig. \ref{fig:Rates}, we have \cite{ng2008queueing}:
\begin{flalign}
\text{Unbounded}&=\lambda-\gamma\nonumber\\
&=\lambda-\lambda P_{\text{rej}}.
\label{eq:relation}
\end{flalign}
On the other hand, the unbind rate, or equivalently, `'Unbounded" shown in Fig. \ref{fig:Rates}, is calculated as below:
\begin{flalign}
\text{Unbounded}=\sum_{n=1}^{N_m}\mu P_n=\mu\left(1-P_0\right),
\label{eq:relation_unbounded}
\end{flalign}
where $P_n$ is the probability of existing $n$ molecules in the system. By equating (\ref{eq:relation}) and (\ref{eq:relation_unbounded}), we have:
\begin{flalign}
\mu\left(1-P_0\right)=\lambda\left(1-P_{\text{rej}}\right).
\label{eq:equalizing}
\end{flalign}
Using the above relation, $P_{\text{rej}}$ is calculated as below:
\begin{flalign}
P_{\text{rej}}=1-\frac{\mu\left(1-P_0\right)}{\lambda}.
\label{eq:p_rej}
\end{flalign}
Now, it is sufficient to obtain $P_0$ and substitute it in the above equation to achieve $P_{\text{rej}}$, and consequently, $\gamma$. To this end, it is necessary to investigate the behavior of a single molecule in a system consisting of a single receptor. In other words, the $M/M/1/1$ queuing system should be considered which is a birth-death process with a single receptor and finite capacity of one molecule. This process is shown in Fig. \ref{fig:single}.
For such a system, the following relation is met:
\begin{flalign}
\lambda P_0=(\mu+\gamma)P_1\Rightarrow P_1=\frac{\lambda}{\mu+\gamma}P_0.
\label{eq:single_relation}
\end{flalign}
The above relation is written according to the balance equation in the queuing theory-based systems. More information about the details of the balance equation can be found in \cite{ng2008queueing}. Moreover, according to $\sum_{n}P_n=1$, we have:
\begin{flalign}
P_0\left[1+\frac{\lambda}{\mu+\gamma}\right]=1.
\label{eq:sum_rule}
\end{flalign}
Therefore, $P_0$ is obtained as below:
\begin{flalign}
P_0=\frac{\mu+\gamma}{\mu+\gamma+\lambda}.
\label{eq:P0single}
\end{flalign}
By inserting the obtained $P_0$ in (\ref{eq:p_rej}), the rejection probability is rewritten as below:
\begin{flalign}
P_{\text{rej}}=\frac{\gamma+\lambda}{\mu+\gamma+\lambda}.
\label{eq:p_rej2}
\end{flalign}
Consequently, the rejection rate would be obtained as below:
\begin{flalign}
\gamma=\lambda P_{\text{rej}}=\frac{\lambda\gamma+\lambda^2}{\mu+\gamma+\lambda}.
\label{eq:gamma_final}
\end{flalign}
In order to obtain the rejection rate, the above equation should be solved; (\ref{eq:gamma_final}) is a quadratic equation (in terms of $\gamma$), the roots of which are as follows:
\begin{flalign}
\gamma^2+\mu\gamma-\lambda^2=0\Rightarrow \gamma_{1,2}=\frac{-\mu\pm\sqrt{\mu^2+4\lambda^2}}{2}.
\label{eq:quadratic}
\end{flalign}
Note that the negative value cannot be assigned to the rejection rate. Therefore, among the obtained solutions, the positive value is selected for the rejection rate, and finally, we have:
\begin{flalign}
\gamma=\frac{1}{2}\left(\sqrt{\mu^2+4\lambda^2}-\mu\right).
\label{eq:gammaaaa}
\end{flalign}
As it can be seen from the above equation, the rejection rate depends on $\mu$ and $\lambda$. According to (\ref{eq:entering_rate}), it can be seen that $\lambda$ is a function of $R$ and $Q$. Therefore, it can be concluded that the rejection rate depends on $R$, $Q$, and also, $\mu$.
\subsubsection{Exit Rate Due to Unbinding Process}
Similar to the previous section, in order to obtain the exit rate, the $M/M/1/1$ queue is considered first. In such a system, the exit rate due to unbinding process is denoted by $\mu$. This value depends on the characteristics of the receptor, and therefore, it would not be changed by varying the parameters $Q$ and $R$. The exit rate due to unbinding process is considered a constant parameter. Note that the exit rate due to the rejection process depends on this constant term, as shown in the previous section.
\subsubsection{Exit Rate in $M/M/N_r/N_m$ System}
Till now, the exit rates $\mu$ and $\gamma$ are obtained by considering the assumption that the system consists of a single receptor, and has a finite capacity of one molecule. However, the capacity of the considered system equals $N_m$. Also, the number of receptors is considered to be $N_r$. It is expected that the exit rate of the molecules from the system is changed with respect to the number of molecules. In other words, the rates are state-dependent. Therefore, it is necessary to consider this parameter in calculating the rates. Each state in Fig. \ref{fig:birth_death} denotes the number of molecules in the reception area, as mentioned before. For instance, the state $N_r$ in Fig. \ref{fig:birth_death}, denotes the case where $N_r$ molecules exist in the reception area. Assume the system is in the state $i$. The death process is a consequence of the unbinding process ($\mu_i$) or the rejection process ($\gamma_i$). Therefore, the rate of the death process (from state $i$ to the state $i-1$) is denoted by $\mu_i+\gamma_i$, as shown in Fig. \ref{fig:birth_death}. For the considered system, we have:
\begin{flalign}
\mu_i=\begin{cases}
\sum_{k=0}^{i}(i-k)\mu  & \quad \text{for } i\leq N_r,\\
\sum_{k=0}^{N_r}(N_r-k)\mu  & \quad \text{for } N_r<i\leq N_m.
\end{cases}
\label{eq:params1}
\end{flalign}
Also, for $\gamma_i$ we have:
\begin{flalign}
\gamma_i=\begin{cases}
\sum_{k=0}^{i}(i-k)\gamma  & \quad \text{for } i\leq N_r,\\
\sum_{k=0}^{N_r}(i-k)\gamma  & \quad \text{for } N_r<i\leq N_m.
\end{cases}
\label{eq:params2}
\end{flalign}
It can be seen that the values of $\mu_i$ and $\gamma_i$ change with respect to the number of molecules, or equivalently, the state of the system. More explanations on how to obtain the above rates are provided in Appendix \ref{Appendix1}.
\subsection{Finite Capacity of the System}
\begin{figure}[t]
	\centering
	\includegraphics[width=0.35\textwidth]{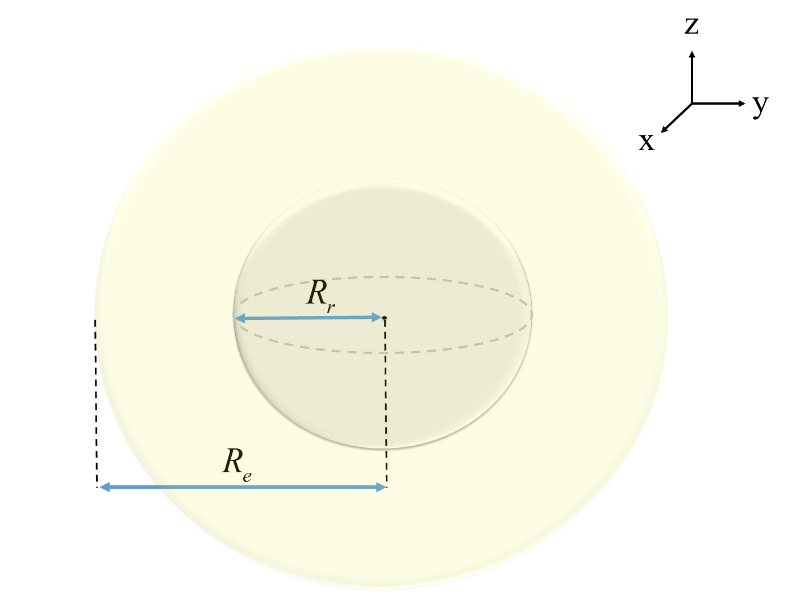}
	\caption{
		Schematic of the reception space around the receiver.}
	\label{fig:area}
\end{figure}

The entering of the molecules to the reception space is modeled by a Poisson distribution, as mentioned before. It is worth noting that the Poisson distribution is the approximation of the Binomial distribution when the number of molecules ($n$) goes to infinity:
\begin{flalign}
\lim\limits_{n \to \infty}\binom{n}{k}\left(\frac{\lambda}{n}\right)^k\left(1-\frac{\lambda}{n}\right)^{n-k}=\frac{e^{-\lambda}\lambda^k}{k!}.
\label{eq:binom}
\end{flalign}
On the other hand, the capacity of the environment is considered to be bounded; consider Fig. \ref{fig:area} where the radius of the receiver is denoted by $R_r$ and the radius of the reception space is denoted by $R_e$. The capacity of the reception area (the yellow environment in Fig. \ref{fig:area}) is obtained as below:
\begin{flalign}
\text{capacity of the reception space}=\frac{4\pi}{3}\nonumber \left(R_e^3-R_r^3\right).
\end{flalign} 
If the radius of the messenger molecule is $R_a$, the maximum number of molecules that can be placed in the reception area would be obtained as below:
\begin{flalign}
\frac{4\pi}{3}\left(R_e^3-R_r^3\right)=N_m\times \frac{4\pi}{3} R_a^3\rightarrow N_m=\lfloor\frac{R_e^3-R_r^3}{R_a^3}\rfloor,
\label{eq:num_molecule}
\end{flalign}
where $\lfloor.\rfloor$ is the lower round operator. It can be concluded that a large number of molecules are released in the environment, which is approximated to be infinite. However, only a finite number of them ($N_m$) can be placed in the reception space. Therefore, the finite number of molecules in the considered system is well justified.\\
\section{Bounds}\label{Bounds}
In drug delivery applications, several drug molecules are transmitted from the transmitter towards the target (receiver). In order to deliver the drug molecules efficiently, the number of active receptors must exceed a specific threshold value. Otherwise, the transmitted drug molecules would not have any effect on the healing process of the disease. However, it should be noted that if the transmission rate of the drug molecules is too high, the system would be saturated and leads to side effects in the human body. In addition, some drugs are expensive, and therefore, using the optimum number of such drug molecules is a matter of concern. It is concluded that a transmission interval should be considered for the drug molecules; so that the transmission rate should not be less than the lowest possible level. Also, it should be adjusted in a way that it does not exceed the highest possible level. In this section, an interval, consisting of a lower and an upper bound, is considered for the transmission rate of the drug molecules.
\subsection{Lower Bound}
Consider the BCSK modulation technique in which the molecules are released in the environment with a constant rate of $Q/\Delta t$, according to (\ref{eq:BCSK}). As discussed before, the concentration of the molecules in steady-state is obtained as below \cite{zhao2020release,bossert1963analysis}:
\begin{flalign}
C=\frac{Q}{4\pi DR\Delta t},
\label{eq:concentration}
\end{flalign}
where $R$ is the distance between the transmitter and the receiver, and $D$ is the diffusion coefficient, as mentioned before. In order to make the drug delivery process efficient, the least number of receptors should be activated. To this end, an occupancy factor $f$ is defined as proportion of receptors that are activated in steady-state. This factor is given by the following equation:
\begin{flalign}
f=\frac{K^+ C}{K^+C+\mu+\gamma},
\label{eq:f_factor}
\end{flalign}
where $K^+$ is the binding constant related to the ligand-receptor complex. This ratio is determined according to the type of drug molecule, and it expresses the minimum required active receptors in order for the delivery process to be effective. According to (\ref{eq:f_factor}), it can be shown that the minimum effective concentration of molecules in order to make the drug delivery process efficient is obtained as below:
\begin{flalign}
C=\frac{f(\mu+\gamma)}{K^+(1-f)}.
\label{eq:least_effective}
\end{flalign}
The above conclusion is obtained using rate theory \cite{paton1961theory}. Using (\ref{eq:concentration}) and (\ref{eq:least_effective}), the lower bound for the transmission rate $Q_{min}$ during the delivery process is obtained as below:
\begin{flalign}
\frac{Q_{min}}{\Delta t}=\frac{4\pi DRf \left(\mu+\gamma\right)}{K^+(1-f)}.
\label{eq:Q_min1}
\end{flalign}
Note that the parameter $\gamma$ is a function of $Q$ which is unknown here. Using (\ref{eq:gammaaaa}) and (\ref{eq:entering_rate}), the above equation is rewritten as below:
\begin{flalign}
\frac{Q_{min}}{\Delta t}=\frac{4\pi DR\mu f(1-f)}{K^+(1-f)^2-f^2/K^+},
\label{eq:Q_min}
\end{flalign}
and the lower bound is obtained accordingly.
This value represents the minimum required concentration of the drug molecules. If the total number of released molecules is lower than the obtained $Q_{min}$, the delivery process would not be efficient.
\subsection{Upper Bound}
In addition to the rate theory, another theory is considered in drug delivery, named occupancy theory \cite{clark1933mode}. This theory reveals that the effectiveness of the drug molecules would be maximized if all of the receptors are activated. In other words, the $f$ ratio should be equivalent to 1. According to (\ref{eq:least_effective}), the concentration of the molecules would be infinite when $f=1$ is considered; an infinite number of molecules should be released in the system to make all of the receptors active. It should be noted that there is a limitation in the number of released molecules in the system; as the number of molecules increases, the number of active receptors would be increased. However, if the number of released molecules exceeds a certain value, side effects in the human body would happen. Therefore, the number of transmitted molecules should not be infinite. The maximum number of molecules that can be placed in the reception space equals $N_m$, as mentioned before. According to this issue, and also according to (\ref{eq:concentration}), the upper bound for the transmission of drug molecules is obtained as below:
\begin{flalign}
\frac{Q}{4\pi DR \Delta t}=N_m\rightarrow \frac{Q_{max}}{\Delta t}=4\pi DRN_m.
\label{eq:Q_max}
\end{flalign}
From (\ref{eq:Q_min}) and (\ref{eq:Q_max}), it can be obtained that the rate of the released molecules should be in the following interval:
\begin{flalign}
Q_{min}\leq Q\leq Q_{max}.
\label{eq:bound}
\end{flalign}
From the above interval, the release dose of the drug molecules is adjusted. The drug dosage lower than the minimum release rate $Q_{min}$ does not affect the target. Also, if the dosage of the drug releasing exceeds the maximum limit $Q_{max}$, it would cause side effects in the body \cite{siepmann2012fundamentals}.
\section{Numerical Results}\label{results}
In this section, the numerical results related to the proposed system are presented. The initial parameters adjusted for simulation are expressed in Table \ref{tab:parameters}. Firstly, the behaviors of the parameters $\lambda$, $\gamma$, and $\mu$ are examined.
\begin{table}[t]
	\caption{Initial parameters of the considered system.}
	\centering
	\begin{tabular}{c|c|c}
		\hline
		Parameter & Variable & Value\\
		\hline
		\hline
		Diffusion coefficient & D & 100 ($\mu m^2/sec$)\\
		Distance between TX and RX & R & 10-20 ($\mu m$)\\
		Number of receptors & $N_r$ & 400-1000\\
		Total number of released molecules & $Q$ & $10^{8}-9.1\times 10^{9}$\\
		Step size & $\Delta t$ & $10^{-4}$ ($sec$)\\
		Ligand-receptor binding constant & $K^+$ & 0.5 ($\mu m^3/sec$)\\
		Unbinding rate & $\mu$ & $10^{-3}$  $(1/\mu sec)$\\
		Radius of RX & $R_r$ & 2 ($nm$)\\
		Radius of the reception space & $R_e$ & 2.3 ($nm$)\\
		Radius of the drug molecule & $R_a$ & 0.01 ($nm$)\\
		\hline
	\end{tabular}
	\label{tab:parameters}
\end{table}%
\begin{figure*}[htbp]
	\centering
	\includegraphics[width=0.8\textwidth]{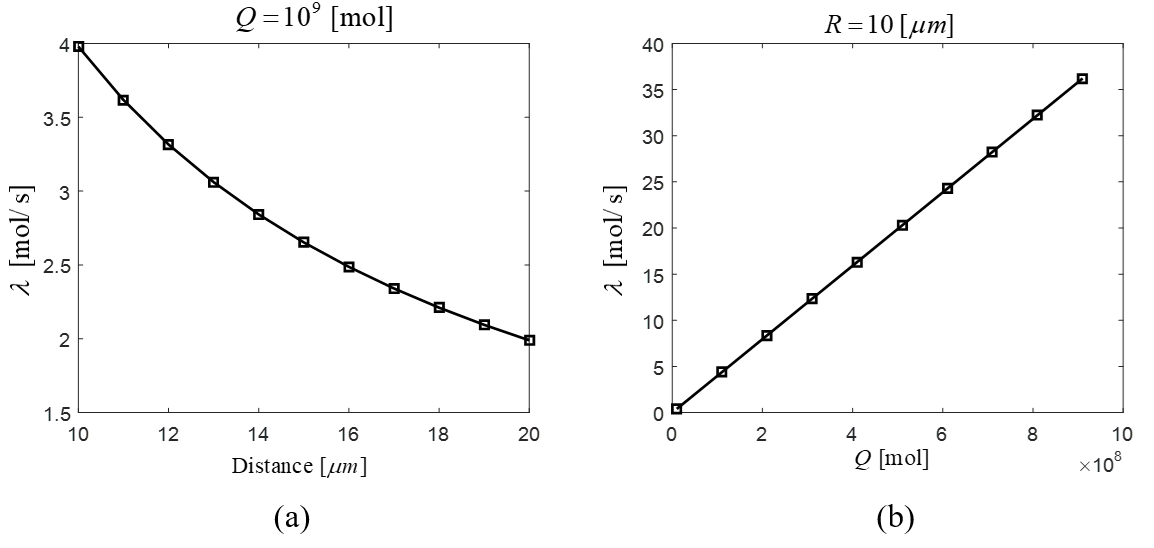}
	\caption{
		Enter rate of the molecules for different values of (a) distance between the transmitter and the receiver, and (b) total number of released molecules.}
	\label{fig:Lambda}
\end{figure*}
\subsection{The Behavior of the Parameters $\lambda$, $\gamma$ and $\mu$}

\begin{figure*}[htbp]
	\centering
	\includegraphics[width=0.8\textwidth]{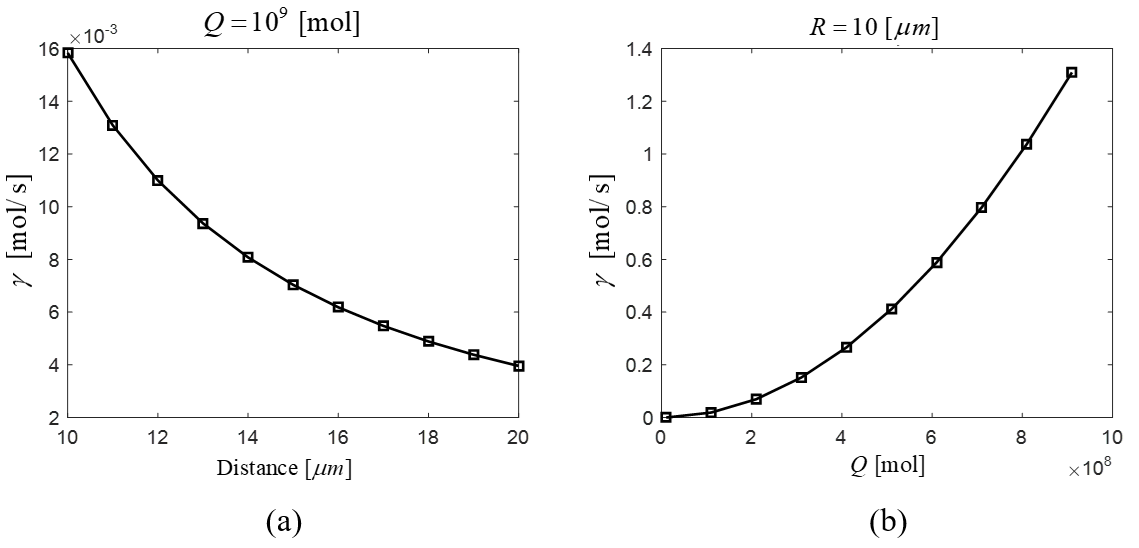}
	\caption{
		Exit rate of molecules due to the rejection process for different values of (a) distance between the transmitter and the receiver, and (b) total number of released molecules.}
	\label{fig:Gamma}
\end{figure*}
\begin{figure*}[htbp]
	\centering
	\includegraphics[width=0.8\textwidth]{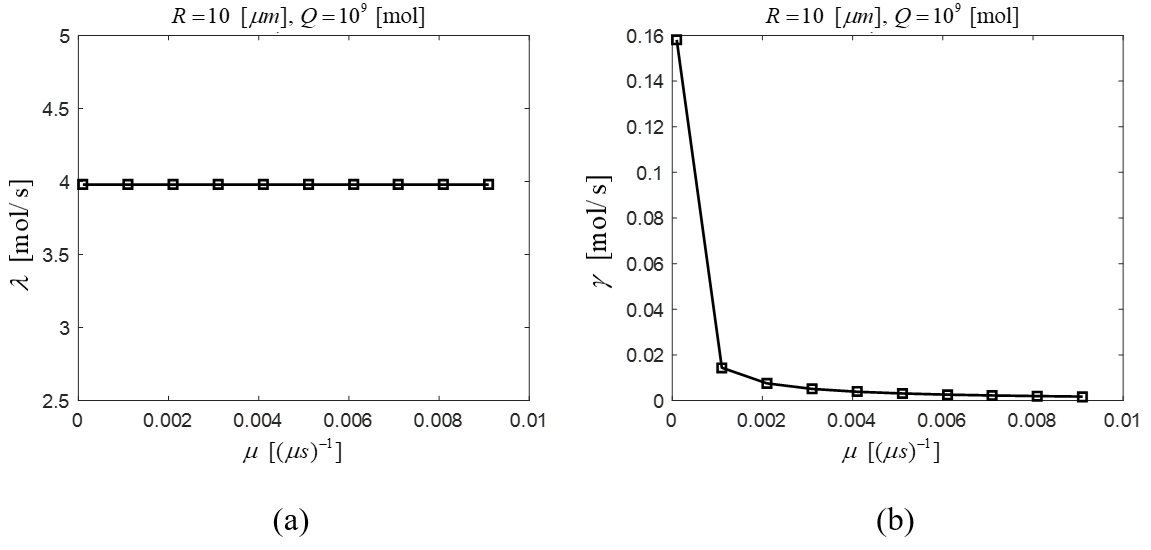}
	\caption{
		(a) Enter rate and (b) rejection rate of the molecules for different values of $\mu$.}
	\label{fig:Mu}
\end{figure*}

Fig. \ref{fig:Lambda} shows the behavior of the enter rate for different values of $R$ and $Q$. Referring to Fig. \ref{fig:Lambda} (a), it can be seen that the enter rate of the molecules is decreased by increasing the distance between the transmitter and the receiver. Also, Fig. \ref{fig:Lambda} (b) shows that the enter rate would be increased by increasing the total number of released molecules. This is also evident from (\ref{eq:entering_rate}).
Fig. \ref{fig:Gamma} shows the rejection rate of the molecules for different values of $R$ and $Q$. It can be seen from Fig. \ref{fig:Gamma} (a) that the value of $\gamma$ decreases by increasing the distance between the transmitter and the receiver. This is due to the fact that as the distance between the transmitter and the receiver increases, the enter rate is decreased. According to (\ref{eq:gammaaaa}), it can be concluded that decreasing the parameter $\lambda$, leads to lower rejection rates. Moreover, from Fig. \ref{fig:Gamma} (b), it can be seen that the rejection rate increases by increasing the total number of released molecules. The reason for this behavior is similar to the one in Fig. \ref{fig:Gamma} (a).
The unbinding rate $\mu$ is a constant parameter that is invariant to $R$ and $Q$, as mentioned before. It is worth investigating the behavior of the parameters $\lambda$ and $\gamma$ for different values of this constant parameter. Considering Fig. \ref{fig:Mu} (a), it can be seen that the enter rate does not depend on the unbinding rate, as expected. This can be concluded from (\ref{eq:entering_rate}). The reason is that the unbinding rate depends on the characteristics of the receptors, while the enter rate of the molecules depends on the system parameters (e.g., $R$ and $Q$), and it does not vary by changing the characteristics of the receptors. Fig. \ref{fig:Mu} (b) shows the dependency of the rejection rate to the unbinding rate; it can be seen that the value of $\gamma$ decreases by increasing the unbinding rate. This is due to the fact that as the parameter $\mu$ increases, the number of molecules that are received by the receptor is increased. In other words, the expected value of the time interval between two consecutive unbinding processes ($1/\mu$) is decreased. Therefore, fewer molecules would leave the system without binding to the receptors. This means that the rejection rate would be decreased.
\begin{figure*}[htbp]
	\centering
	\includegraphics[width=0.8\textwidth]{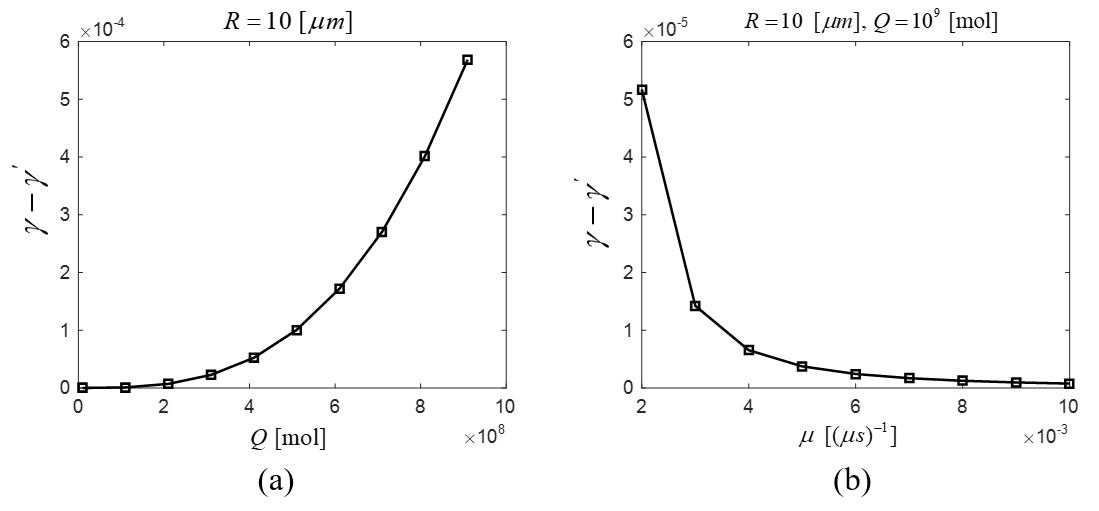}
	\caption{
		The difference between the obtained $\gamma$ and $\gamma^{\prime}$ for different values of (a) $Q$ and (b) $\mu$.}
	\label{fig:Comparison}
\end{figure*}
\begin{figure*}[htbp]
	\centering
	\includegraphics[width=0.8\textwidth]{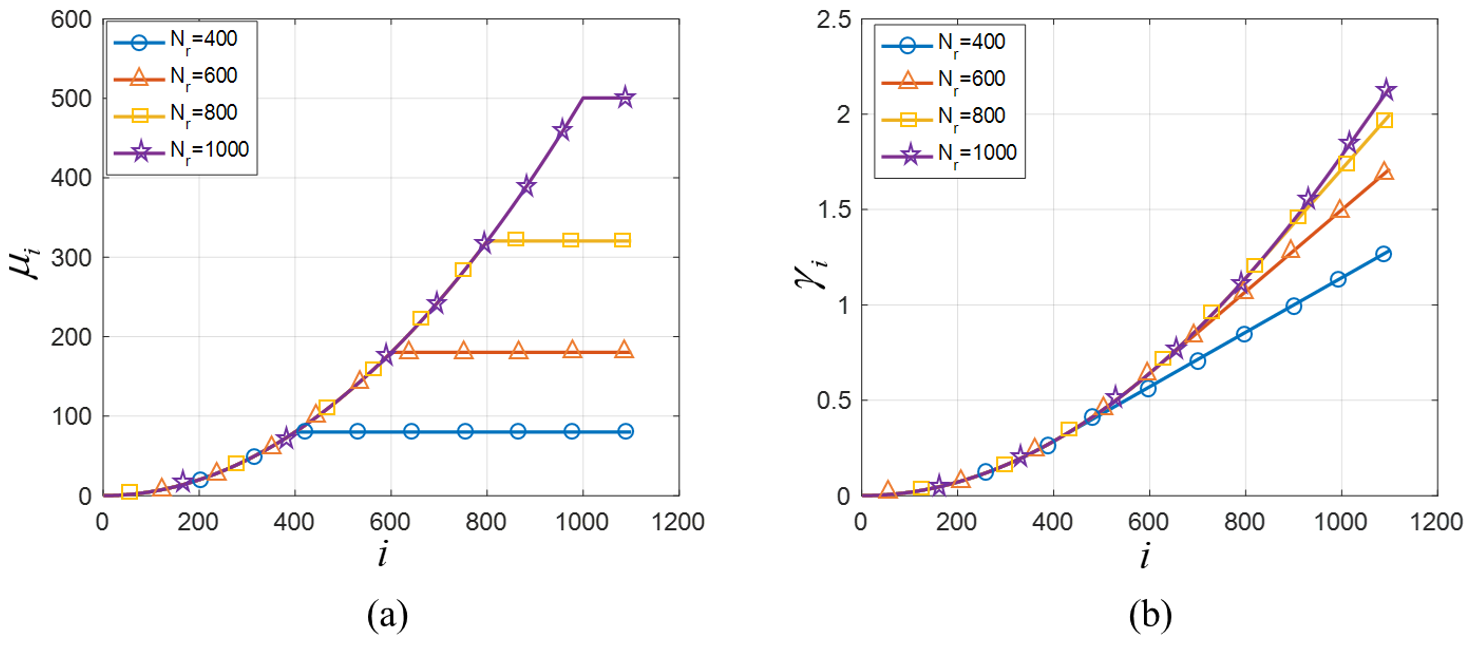}
	\caption{The behavior of (a) $\mu_i$ and (b) $\gamma_i$ for different values of $N_r$ .}
	\label{fig:Different_Nr}
\end{figure*}
\subsection{Comparison Between Parameters $\gamma$ and $\gamma^{\prime}$}
The difference between the rejection rate obtained in \cite{felicetti2016simple}, which we show hereafter as $\gamma^{\prime}$, and the rejection rate obtained in this study is discussed before. Note that $\gamma^{\prime}=\frac{\lambda^2}{\mu+\lambda}$. It is worth comparing these two parameters in order to see if there is a difference between them. Fig. \ref{fig:Comparison} shows the difference between these two parameters for different values of $Q$ and $\mu$. It can be seen from the figure that the parameters $\gamma$ and $\gamma^{\prime}$ are not equivalent; in the proposed system, the rejection rate $\gamma$ is not only influenced by active receptors. In addition to active receptors, the rejection process can occur due to some other reasons such as random movement of the molecules. Therefore, it is expected that $\gamma\geq\gamma^{\prime}$ in all situations. Fig. \ref{fig:Comparison} (a) shows that the difference between $\gamma$ and $\gamma^{\prime}$ is increased as the total number of released molecules increases. This means that considering the proposed system model, it can be seen that more molecules would be rejected from the system compared to the model presented in \cite{felicetti2016simple}. Moreover, it can be observed from Fig. \ref{fig:Comparison} (b) that the obtained $\gamma$ would reach the parameter $\gamma^{\prime}$ as the unbinding rate $\mu$ increases; as the parameter $\mu$ increases, more molecules would be caught by the receptors, as concluded from Fig. {\ref{fig:Mu} (b)}. Therefore, fewer molecules would be rejected (especially, due to other reasons such as random movement of the molecules). Therefore, $\gamma\approx\gamma^{\prime}$ as the parameter $\mu$ increases.
\begin{figure*}[htbp]
	\centering
	\includegraphics[width=1\textwidth]{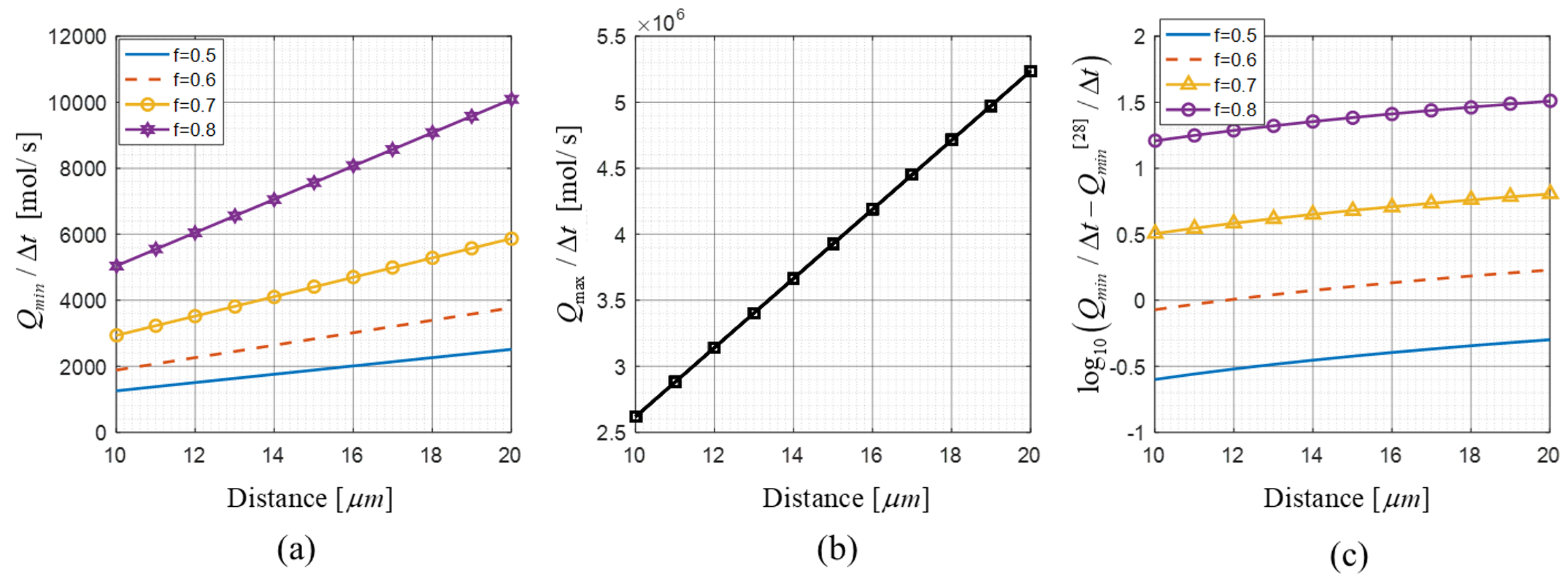}
	\caption{
		(a) $Q_{min}/\Delta t$ and (b) $Q_{max}/\Delta t$ v.s different values of distance between the transmitter and the receiver, (c) the difference between the proposed minimum release rate and the one obtained in [28].}
	\label{fig:Interval}
\end{figure*}
\subsection{The Behavior of the Parameters $\mu_i$ and $\gamma_i$}
The parameter $\gamma$ is obtained from (\ref{eq:gammaaaa}), and its behavior is investigated v.s different values of $\mu$ and $Q$ in Fig. \ref{fig:Gamma} for a system consisting of a single receptor and one molecule. However, it is assumed that the receiver consists of $N_r$ number of receptors. Moreover, the number of drug molecules in the system is assumed to be $\leq N_m$, as mentioned before. In such a case, it is expected that the number of receptors affects the value of the parameters $\mu$ and $\gamma$, as shown in (\ref{eq:params1}) and (\ref{eq:params2}). Fig. \ref{fig:Different_Nr} shows the behavior of these parameters for different values of $N_r$. From Fig. \ref{fig:Different_Nr} (a), it can be seen that the value of the parameter $\mu_i$ increases by increasing the number of receptors. More precisely, the expected value of the unbinding process $1/\mu_i$ is decreased as the number of receptor increases. Therefore, it can be concluded that increasing the number of receptors leads to more molecules be caught by the receiver, as expected. Note that the parameter $\mu_i$ increases until the number of molecules is less than (or equal to) the number of receptors. As the number of molecules in the system equals the number of receptors, the parameter $\mu_i$ would be constant; i.e. $\mu_i=\mu_{const}$ for $i>N_r$, as shown in (\ref{eq:params1}). From Fig. \ref{fig:Different_Nr} (b), it can be concluded that for different values of $N_r$, the rejection rate would be increased as the number of molecules in the system increases. Also, it can be seen that for different values of $N_r$, the rejection rate is not changed until the number of molecules in the system reaches a threshold (approximately 550). Note that as the number of receptors increases, increasing the rejection rate (by increasing the number of molecules) slows down.
\subsection{The Lower and Upper bound for Releasing The Molecules}
An interval, consisting of a maximum and minimum allowable dosage for releasing the drug molecules is defined according to (\ref{eq:Q_min}) and (\ref{eq:Q_max}), respectively. Referring to the defined interval, it can be seen that the lower bound depends on the type of drug molecules. In other words, it depends on the $f$ ratio. However, according to (\ref{eq:Q_max}), it can be seen that the upper bound is
independent of this factor. Fig. \ref{fig:Interval} shows the defined interval for different values of $R$. The lower bound for different values of $f$ ratio is depicted in Fig. \ref{fig:Interval} (a). Also, the upper bound for the number of released molecules is presented in Fig. \ref{fig:Interval} (b). It can be seen from the figure that as the distance between the transmitter and the receiver increases, the lower and upper bounds of the transmitted drug molecules are increased, as expected. Note that there is a linear relationship between the minimum releasing rate and distance, as shown in Fig. \ref{fig:Interval} (a). This issue is obvious from (\ref{eq:Q_min}).\\
In \cite{zhao2020release}, the minimum release rate is obtained for the corresponding system model in which the rejection rate is not considered. It is worth comparing the minimum number of released molecules obtained in this study with the one obtained in \cite{zhao2020release}. The logarithmic graph  of the difference between these two parameters can be seen from Fig. \ref{fig:Interval} (c). The minimum release rate obtained from \cite{zhao2020release} is shown as $Q^{[28]}_{min}$ here. It can be seen that the lower bound corresponding to the proposed system model results in a higher value of released molecules, as expected; according to the proposed reception model, some of the drug molecules in the system would leave the system without any binding process. Due to the so-called rejected molecules, more molecules are required to be released toward the target in order to reach the minimum acceptable active receptors. Note that as the parameter $f$ increases, the length of the interval would be smaller. This issue is due to the fact that as the parameter $f$ increases, more receptors should be activated in order to make the drug delivery process efficient. As a result, the minimum number of transmitted molecules would be increased. This can be also concluded from Fig. \ref{fig:Interval} (a). Also, note that increasing the value of $f$ ratio, the difference between the proposed $Q_{min}$ and $Q^{[28]}_{min}$ would be increased, as shown in Fig. \ref{fig:Interval} (c).
\begin{figure*}[htbp]
	\centering
	\includegraphics[width=0.9\textwidth]{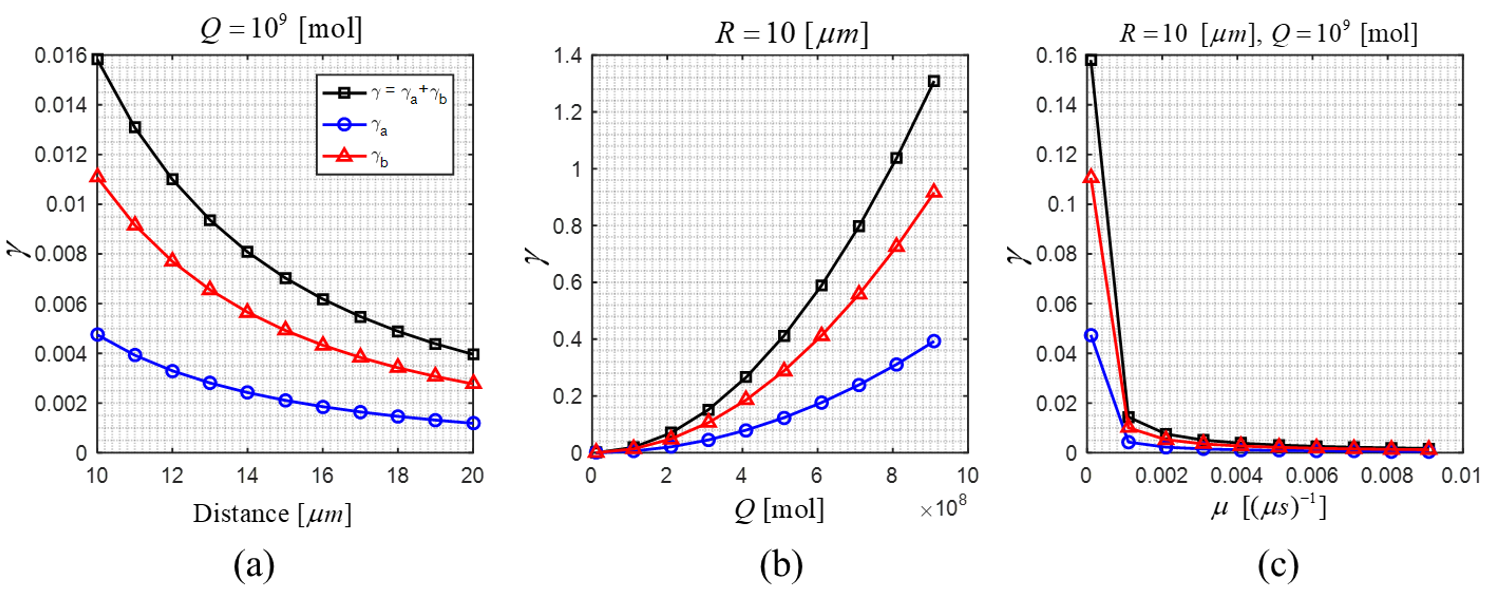}
	\caption{The obtained values of $\gamma$, $\gamma_a$ and $\gamma_b$ separately v.s different values of (a) $R$, (b) $Q$ and (c) $\mu$ for $\alpha=0.3$.}
	\label{fig:Example_gamma}
\end{figure*}
\begin{figure*}[htbp]
	\centering
	\includegraphics[width=0.9\textwidth]{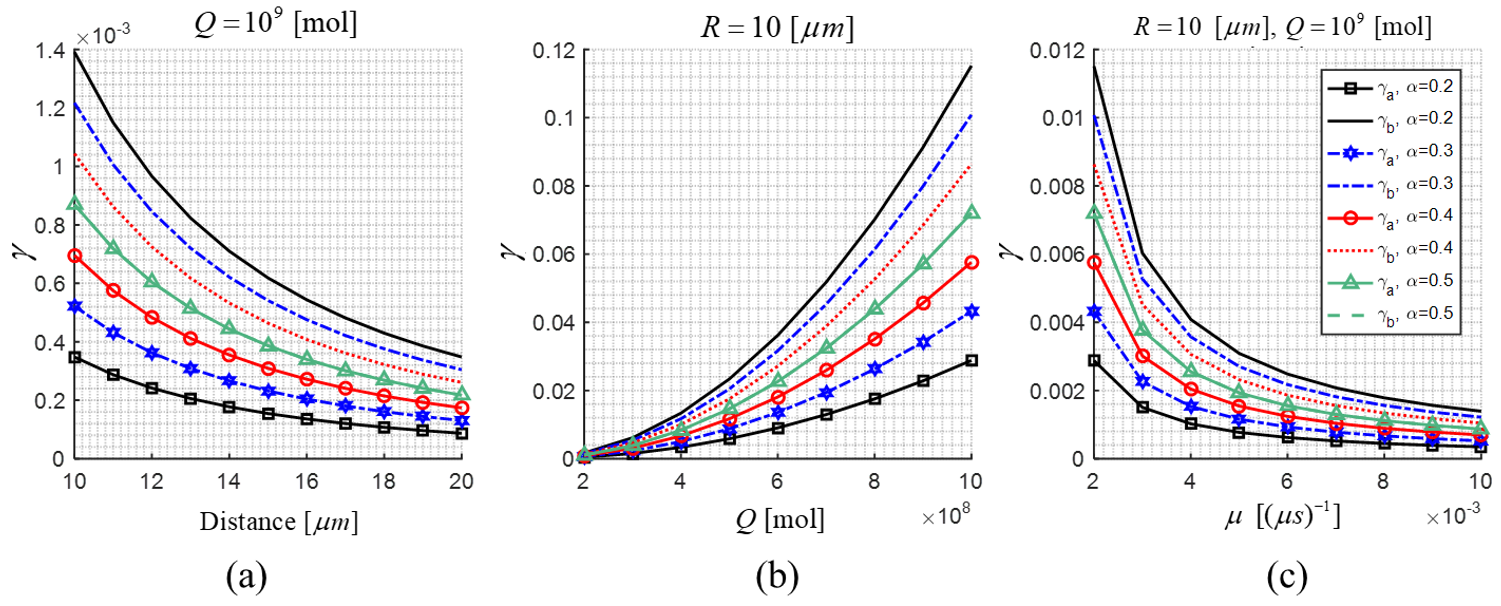}
	\caption{The obtained values of $\gamma_a$ and $\gamma_b$ v.s different values of (a) $R$, (b) $Q$ and (c) $\mu$ for different $\alpha$s.}
	\label{fig:Gamma_divided}
\end{figure*}
\subsection{Detailed Analysis of Parameters $\gamma$}
The rejection rate obtained in this study is divided into two parts: rejection due to active receptors ($\gamma_a$), and rejection due to some other reasons, such as random movements of the molecules ($\gamma_b$), as shown in Fig. \ref{fig:Rates}. In other words, the obtained rejection rate could be written as below:
\begin{flalign}
\gamma=\gamma_a+\gamma_b,
\label{eq:alanyse}
\end{flalign}
where:
\[\begin{cases}
\gamma_a&=\alpha \gamma,\\
\gamma_b&=(1-\alpha) \gamma,
\end{cases}
\]
where $\alpha\leq1$ is a constant parameter. Fig. \ref{fig:Example_gamma} shows the role of each part ($\gamma_a$ and $\gamma_b$) in the construction of $\gamma$ for different values of $R$, $Q$ and $\mu$. In this figure, it is assumed that $\alpha=0.3$. It can be seen that in the considered case, the rejection rate due to active receptors plays a less important role in the final rejection rate. In other words, the rejection process is mainly influenced by the random movement of the molecules. It should be noted that as the parameter $\alpha$ changes, the participation rate of each part $\gamma_a$ and $\gamma_b$ would change. Fig. \ref{fig:Gamma_divided} shows the behavior of each part for different values of $\alpha$. The parameter $\alpha$ is interpreted as the probability of the molecules being rejected due to active receptors. Similarly, $1-\alpha$ is equivalent to the probability of the molecules being rejected due to other reasons (e.g., random movement of the molecules). Fig. \ref{fig:Gamma_divided} shows that for $\alpha=0.5$, the level of participation of each part $\gamma_a$ and $\gamma_b$ would be equal, as expected.
\section{Conclusion}\label{conclusion}
In this paper, a reception model was proposed for drug delivery application using queuing theory. In contrast to previous reception models, in the proposed model the rejection rate of the molecules was considered to be not only due to active receptors, but also from other reasons, such as random movement of the molecules. Moreover, the rejection rate was investigated by the exit rate of the molecules. In other words, in the proposed model, the arrival rate of the molecules to the system was considered to be constant. In order to model the reception process, a system consisting of a single receptor and one molecule was investigated first. Then, the result was generalized to the $M/M/N_r/N_m$ queuing system. When the molecules enter the system, different scenarios could happen; each of them can be received by the receptor or leave the system without any binding process. Leaving the molecules, or equivalently, the rejection process could be due to active receptors or random movement of the molecules. The rate of each scenario was discussed in this study and their behavior was analyzed numerically for different values of the system parameters. Finally, in order to define a range of allowable dosage of the drug molecules and manage the drug delivery process, an interval, consisting of lower and upper bounds was presented.
\appendices

\section{More Explanations about Obtaining the Parameters $\mu_i$ and $\gamma_i$}\label{Appendix1}
The parameters $\mu_i$ and $\gamma_i$ denote the unbinding rate and the rejection rate of the molecule, respectively. It should be noted that the unbinding and rejection probabilities of the molecules is obtained by multiplying the corresponding rates with $\Delta t$, which is defined as the time duration in which the state of the system is changed.\\
\indent In particular, consider the probability of unbinding a molecule as $\mu\Delta t$. The time duration that the process goes from state $n$ to the state $n-1$ is modeled using the exponential distribution as $\mu e^{-\mu t}$. It can be found that the probability that the state is changed during the interval $\left[0, \Delta t\right]$ is obtained as below:
\begin{flalign}
\int_{0}^{\Delta t}\mu e^{-\mu t}dt=1-e^{-\mu\Delta t}=\mu \Delta t+o(\Delta t).
\label{eq:1}
\end{flalign}
Note that the right-hand side of the above equation is written according to the Taylor series. The probability that the state is not changed during this interval is obtained as below:
\begin{flalign}
P(\text{the state is not changed})=1-\mu\Delta t+o(\Delta t).
\label{eq:2}
\end{flalign}
Note that when $n$ molecules exist in the system ($n\leq N_r$), $n+1$ different scenarios could be imagined:
\begin{itemize}
	\item Scenario 1: $n$ number of receptors are activated,
	\item Scenario 2: $n-1$ number of receptors are activated,\\
	$\vdots$
	\item Scenario n: a single receptor is activated,
	\item Scenario n+1: none of the receptors are activated,
\end{itemize}
Consider the first scenario; the case in which $n$ parallel and independent receptors are activated. The probability that none of the receptors are activated is equivalent to the probability that no unbinding process occurs. This probability is obtained according to (\ref{eq:2}) as below:
\begin{flalign}
&P(\text{none of the n receptors is unbounded})\nonumber\\
&=\left[1-\mu\Delta t+o(\Delta t)\right]^n,
\label{eq:3}
\end{flalign}
and consequently, the probability that at least one receptor is unbounded would be achieved as below:
\begin{flalign}
&P(\text{at least one receptor is unbounded})\nonumber\\
&=1-\left[1-\mu\Delta t+o(\Delta t)\right]^n=n\mu\Delta t+o(\Delta t).
\label{eq:4}
\end{flalign}
Note that $o(\Delta t)$ is a small value ($\approx 0$). Also, note that the considered $\Delta t$ is small enough, and therefore, the probability that at least one receptor is unbounded, equals to the probability of unbinding exactly one receptor, which is equivalent to $n\mu$.\\
The above explanation is performed based on the first scenario ($n$ active receptors). The other scenarios are similar to the above explanation, except that the number of active receptors is considered instead of $n$. Finally, the summation of the rates obtained from each scenario results the final exit rate due to unbinding process. Therefore, (\ref{eq:params1}) is concluded. Similar explanations are also true for the rejection rate, and therefore, (\ref{eq:params2}) is concluded.

\bibliographystyle{ieeetr}
\bibliography{Ref}

\end{document}